\documentclass[%
 reprint,
superscriptaddress,
 amsmath,amssymb,
 aps,
]{revtex4-2}
\usepackage{xcolor}
\definecolor{apsblue}{RGB}{51,51,153}
\usepackage{natbib}

\usepackage[
    colorlinks=true,
    linkcolor=apsblue,   
    citecolor=apsblue,   
    urlcolor=apsblue     
]{hyperref}

\makeatletter
\def\NAT@spacechar{} 
\makeatother

\usepackage{physics,amsmath,amssymb,theorem,color,epsfig,epic,subfigure,hhline,graphicx}
\usepackage{dcolumn}
\usepackage{bm}

\begin{document}

\preprint{APS/123-QED}

\title{Optimal Gaussian networks for private distributed quantum sensing
}

\author{Hanbom Yoo}%
\thanks{These authors contributed equally to this work.}
\affiliation{%
 Department of Physics, Yonsei University, Seoul 03722, Republic of Korea}
 \author{Byeongyun Yang}%
 \thanks{These authors contributed equally to this work.}
 \affiliation{%
 Department of Physics, Yonsei University, Seoul 03722, Republic of Korea}
  \author{Hyunwoo Yoo}%
 \affiliation{%
 Department of Physics, Yonsei University, Seoul 03722, Republic of Korea}
\author{Seongjin Hong}%
 \email{shong@yonsei.ac.kr}
\affiliation{%
 Department of Physics, Yonsei University, Seoul 03722, Republic of Korea}%
 \affiliation{%
 Department of Quantum Information, Yonsei University, Seoul 03722, Republic of Korea}

\date{\today}

\begin{abstract}

Private distributed quantum sensing aims to estimate an authorized collective parameter while preventing independent estimation of individual local parameters. Here, we analytically characterize Gaussian quantum networks satisfying this perfect local privacy condition. Using a graph representation of the Gaussian pairing matrix, we show that two-mode squeezed vacuum states constitute the essential building blocks of privacy-preserving Gaussian states. We then analytically derive the optimal sensitivity under perfect local privacy and determine a Gaussian probe that achieves Heisenberg scaling with respect to both the photon number and the number of sensing modes. Using local homodyne measurements and maximum-likelihood estimation, we numerically verify quantum-enhanced sensitivity beyond the shot-noise limit while preserving perfect local privacy. We further show that the local-privacy condition is preserved under arbitrary phase-independent quantum channels, including optical loss. Our results provide a general framework for constructing and optimizing privacy-preserving continuous-variable quantum sensing networks.
\end{abstract}

\maketitle

Quantum metrology enables sensitivities beyond classical limits by leveraging quantum features such as squeezing and entanglement \cite{giova11,piran18,polin20,giova04,lawri19,lawri19}. Recently, distributed quantum sensing (DQS) has attracted intense interest owing to its capability to achieve Heisenberg scaling (HS) with respect to both the photon number and the number of sensing modes for estimating global properties of spatially distributed parameters \cite{zhang21,gessn20,oh20,oh22,humphreys13,ge25,kwon19,gessn18,pezze25,kwon22,kim25,kim24,hong21,hong25,du25,liu21}. In particular, recent progress in continuous-variable platforms has demonstrated the potential of Gaussian quantum networks for scalable DQS \cite{zhang21,gessn20,oh20,oh22,matsu19,proct18,kwon22,guo20}, owing to their deterministic generation \cite{lars19,takeda19,yonezu23}, compatibility with existing photonic technologies \cite{enomoto21}, scalability to large quantum networks \cite{lars19,asava19,yoko13}, and compact descriptions in terms of covariance matrices \cite{weed12,serafini23} and graph representations \cite{menicucc12,asava22,cardin24,wang20}.

An important emerging question complementary to precision enhancement concerns privacy in distributed sensing networks \cite{hassani25,de26,bugalho25,ho26,farokhi26,kianvash26,bianchi26,li26,yang24,wang26,spencer26,sh22}. Privacy in DQS aims to estimate an authorized collective parameter while restricting access to other encoded parameter information. A stringent notion of perfect privacy requires that the encoded state contain information only about the target parameter. This condition is equivalent to requiring a rank-one quantum Fisher information matrix (QFIM) whose support is aligned with the target direction \cite{hassani25,bugalho25}. Such perfect privacy can be realized with appropriately designed discrete-variable states, including GHZ-type probes \cite{hassani25,bugalho25,ho26,sh22}. For Gaussian quantum networks, however, finite-energy states generally retain information about additional collective parameter combinations, so the rank-one condition can only be approached asymptotically \cite{alushi26,pereira26}.

In parallel, a more operational notion of privacy has been developed, in which individual local parameters remain inaccessible even when additional collective information is retained \cite{de26,namkung26,li26}. Here, we define perfect local privacy as the condition under which no individual local parameter is independently estimable, while the authorized collective parameter remains estimable. This notion is particularly relevant to distributed sensing, where protecting each local parameter can be sufficient while still allowing useful global information to be extracted \cite{de26,namkung26}. The central question is then which sensing resources can simultaneously preserve this privacy condition and provide quantum-enhanced sensitivity beyond the shot-noise limit (SNL) for the authorized collective parameter. Although several specific probe states and sensing configurations have been studied \cite{de26,alushi26,ho26,sh22}, a general characterization of privacy-preserving Gaussian networks and their optimal sensing sensitivity has not yet been established.

In this Letter, we analytically characterize the general structure of Gaussian quantum networks that satisfy perfect local privacy in DQS. Starting from an arbitrary Gaussian state, we analytically characterize the full class of privacy-preserving probe states and introduce a graph representation that provides an intuitive description of the allowed network structures. We then derive the optimal sensitivity achievable within this class and determine the corresponding optimal Gaussian probe state. We show that the optimal probe achieves HS with respect to both the photon number and the number of sensing modes while preserving perfect local privacy. Finally, we propose an experimentally feasible implementation based on the analytically derived optimal Gaussian probe together with passive linear optics and local homodyne measurements, and validate both privacy and quantum-enhanced sensitivity for the authorized collective parameter through maximum-likelihood simulations. We further show that the local-privacy condition is preserved under phase-independent quantum channels, including optical loss, demonstrating the robustness of the proposed framework against realistic imperfections.

\begin{figure}[t]
\centering\includegraphics[width=\columnwidth]{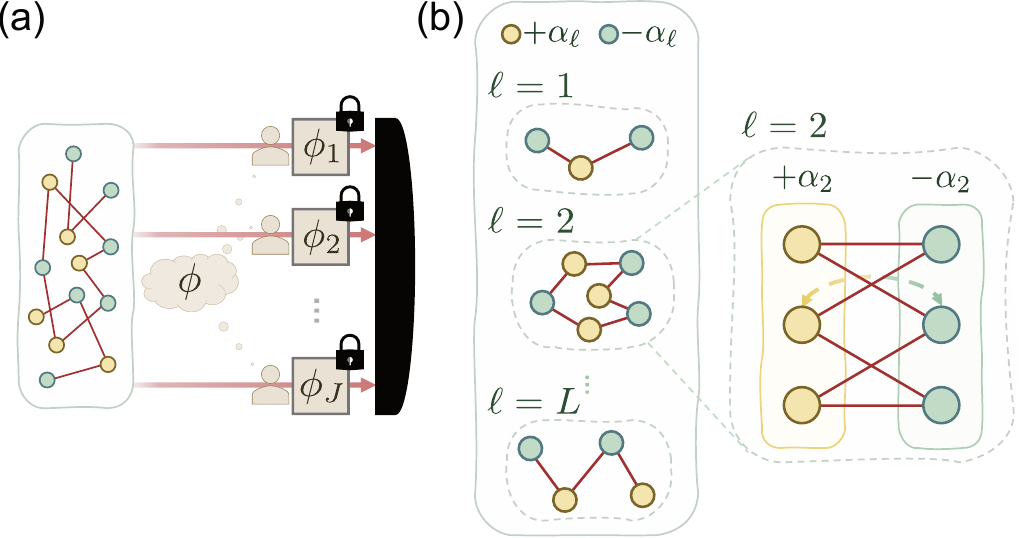}
    \caption{Schematic of private DQS. (a) A multimode quantum probe interrogates spatially separated phase shifts \(\{\phi_j\}_{j=1}^{J}\), while the sensing network is designed such that the target parameter $\phi=\sum_{j=1}^{J}w_j\phi_j$ is accessible and all protected individual local parameters remain hidden. (b) After a suitable permutation, nodes are grouped into independent $\ell$-block, each partitioned into $+\alpha_\ell$ and $-\alpha_\ell$ sectors again. The enlarged $\ell=2$ block illustrates the corresponding connectivity between the two sectors and non-connectivity within a sector.
}
\label{fig1}
    \end{figure}

We consider the private distributed phase-sensing protocol, as shown in Fig.~\ref{fig1}(a), where the goal is to estimate the global parameter
$\phi = \mathbf{w}^{\mathsf{T}}\boldsymbol{\phi} = \sum_{j=1}^{J} w_j \phi_j$,
a weighted combination of $J$ local phase shifts
$\boldsymbol{\phi} = (\phi_1, \ldots, \phi_J)^{\mathsf{T}}$
with weight vector $\mathbf{w} = (w_1, \ldots, w_J)^{\mathsf{T}}$ while keeping the individual local parameters hidden. Each phase shift is encoded onto a probe state via the unitary
$\hat{U}_{\boldsymbol{\phi}} = \exp(-i \sum_{j=1}^J \phi_j \hat{a}_j^{\dagger} \hat{a}_j)$. 
For the estimation of $\phi = \mathbf{w}^{\mathsf{T}}\boldsymbol{\phi}$,
the mean-squared error (MSE) of any unbiased estimator satisfies \cite{li12,kwon25}
\begin{equation}
    \Delta^2\phi \ge \mathbf{w}^{\mathsf{T}}\boldsymbol{\mathrm{F}}^{+}\mathbf{w}
    \ge \mathbf{w}^{\mathsf{T}}\boldsymbol{\mathrm{H}}^{+}\mathbf{w},
    \label{eq:bound}
\end{equation}
where $\boldsymbol{\mathrm{F}}$ is the classical Fisher information matrix (CFIM)
with elements
$\mathrm{F}_{ij} = \sum_{\boldsymbol{x}}
\partial_{\phi_i} p(\boldsymbol{x}|\boldsymbol{\phi})\,
\partial_{\phi_j} p(\boldsymbol{x}|\boldsymbol{\phi})\,
/ p(\boldsymbol{x}|\boldsymbol{\phi})$~\cite{paris09},
where $p(\boldsymbol{x}|\boldsymbol{\phi})
= \mathrm{Tr}[\hat{\rho}_{\boldsymbol{\phi}}\hat{\Pi}_{\boldsymbol{x}}]$
is the Born-rule probability associated with the POVM $\{\hat{\Pi}_{\boldsymbol{x}}\}$,
and $\boldsymbol{\mathrm{H}}$ is the QFIM, whose
elements are defined by
$H_{ij}=\frac{1}{2}\mathrm{Tr}
[\hat\rho_{\boldsymbol{\phi}}(\hat L_i\hat L_j+\hat L_j\hat L_i)]$~\cite{liu20},
where the symmetric logarithmic derivatives $\hat L_i$ satisfy
$\partial_{\phi_i}\hat\rho_{\boldsymbol{\phi}}
=(\hat L_i\hat\rho_{\boldsymbol{\phi}}
+\hat\rho_{\boldsymbol{\phi}}\hat L_i)/2$.
Here, $+$ denotes the Moore--Penrose pseudoinverse \cite{barata12,namkung24}. Since the local parameters are not independently estimable in private DQS, the CFIM and QFIM are generally singular \cite{kwon25}; the pseudoinverse nevertheless yields a well-defined sensitivity bound for the global parameter $\phi$ \cite{kwon25,namkung24}.

Here, perfect local privacy is defined as the condition under which no individual local parameter admits a finite-variance unbiased estimator, while the authorized collective parameter remains estimable. An equivalent criterion can be expressed directly in terms of the derivatives of the encoded probe state \cite{kwon25,de26}. Specifically, perfect local privacy holds if and only if there exists a vector $\boldsymbol{\alpha}=(\alpha_1,\ldots,\alpha_J)^T$, with $\alpha_j\in\mathbb{C}\setminus{0}$ for all $j$, such that
\begin{equation}
\sum_{j=1}^{J}\alpha_j
\frac{\partial \hat{\rho}_{\boldsymbol{\phi}}}{\partial \phi_j}
=0,
\label{eq:priv}
\end{equation}
where $\hat{\rho}_{\boldsymbol{\phi}}
=\hat{U}_{\boldsymbol{\phi}}\hat{\rho}
\hat{U}_{\boldsymbol{\phi}}^\dagger$ denotes the probe state after phase encoding. Here, for the $j$-th basis unit vector $\bm e_j:=(0,\ldots,0,1,0,\ldots,0)$, the derivative $\partial\hat \rho _{\bm \phi}/\partial\phi_j=\lim_{\epsilon \to 0}(\hat\rho_{\bm\phi+\epsilon \bm e_j}-\hat\rho_{\bm\phi})/\epsilon$ is understood in the trace-norm sense, namely, $\lim_{\epsilon \to 0}X_\epsilon=X$ means $\lim_{\epsilon \to 0}||X_\epsilon-X||_1=0$, where $||\cdot||_1:=\operatorname{Tr}\sqrt{(\cdot)(\cdot)^\dagger}$ denotes the trace norm.

The underlying intuition is that the partial derivatives $\partial\hat{\rho}_{\boldsymbol{\phi}}/\partial\phi_j$ are linearly dependent, so that a change in any individual local parameter can be compensated by changes in the remaining parameters. As a result, the local parameters cannot be independently distinguished from the encoded state. In terms of the QFIM, $\mathbf H$, Eq.~\eqref{eq:priv} means that there exists a nonzero direction $\boldsymbol{\alpha}$ along which the encoded state carries no information, i.e., $\boldsymbol{\alpha}\in\operatorname{ker}\mathbf H$. In contrast, the target parameter $\phi=\mathbf w^T\boldsymbol{\phi}$ is estimable only if its weight vector $\mathbf w$ lies within the information-carrying subspace of the QFIM, $\mathbf w\in\operatorname{supp}\mathbf H$ \cite{li12,namkung24,kwon25}. Since the kernel and support of the positive-semidefinite QFIM are orthogonal, the privacy direction must therefore satisfy $\boldsymbol{\alpha}\perp\mathbf w$.

We first consider pure Gaussian states, which provide a clear characterization of private DQS, while the convexity of the QFIM ensures that the optimal sensitivity under perfect local privacy can be attained by a pure state \cite{liu20}. The extension to general Gaussian states, including mixed states, is provided in the Supplemental Material (SM) Sec.~G. Any pure Gaussian state can be generated by single-mode squeezers, displacement operations, and passive interferometers according to the Bloch--Messiah decomposition \cite{braunstein05},
$|\psi\rangle = \hat{U}_K \hat{S} (\boldsymbol{r}) \hat{ \mathcal D} (\bm \beta)|0\rangle$,
where $\hat{ \mathcal D}  (\bm \beta):=\bigotimes_{j=1}^J \hat{ \mathcal D} _j(\beta_j)$ and $\hat{S}(\boldsymbol{r}) := \bigotimes_{j=1}^J \hat{S}_j(r_j)$ are the product of displacement operators and single-mode squeezing operators, respectively, 
and $\hat{U}_K$ is a passive interferometer represented by $\mathrm U_K \in U(J)$ via
$\hat{U}_K \hat{\boldsymbol{a}}^\dagger \hat{U}_K^\dagger = \mathrm U_K^{\mathsf{T}} \hat{\boldsymbol{a}}^\dagger$ \cite{weed12}. A nonzero local displacement carries information about the corresponding local phase and is therefore incompatible with perfect local privacy~\cite{pinel13}; hence, $\boldsymbol\beta=0$ (see SM Sec.~A). Using the Fock-basis representation $\hat{S}(\boldsymbol{r})|0\rangle=\prod_{j=1}^J\left[(\cosh r_j)^{-1/2} \exp \left( \frac{1}{2}\tanh r_j \hat {a}_j^{\dagger2} \right)\right]|0\rangle$ \cite{cariolaro15},
and applying $\hat{U}_K$, the state can be written as \cite{hamilton17,deshpande22,yu23,banchi20,zhang25}
\begin{align}
|\psi\rangle
&= \left[\det(\cosh \mathrm R)\right]^{-1/2}
\exp\!\left(
\frac{1}{2} \hat{\boldsymbol{a}}^{\dagger\mathsf{T}} \mathrm U_K \tanh \mathrm R\, \mathrm U_K^{\mathsf{T}} \hat{\boldsymbol{a}}^\dagger
\right)|0\rangle
\nonumber\\
&=: \mathcal{N}_Z
\exp\!\left(
\frac{1}{2} \hat{\boldsymbol{a}}^{\dagger\mathsf{T}} \mathrm Z \hat{\boldsymbol{a}}^\dagger
\right)|0\rangle,
\label{eq:stateafterUK}
\end{align}
where $\mathrm R := \mathrm{diag}(r_1,\ldots,r_J)$,
$\mathrm Z := \mathrm U_K \tanh \mathrm R\, \mathrm U_K^{\mathsf{T}}$,
and $\mathcal{N}_Z = \det(\mathrm I_J - \mathrm Z\mathrm Z^\dagger)^{1/4}$. We refer to $\mathrm Z$ as the pairing matrix \cite{cariolaro15}, since
\begin{equation}
\frac{1}{2}\hat{\boldsymbol a}^{\dagger\mathsf T}
\mathrm Z
\hat{\boldsymbol a}^{\dagger}
=
\frac{1}{2}\sum_{j,k}
Z_{jk}\hat a_j^\dagger \hat a_k^\dagger,
\end{equation}
where $Z_{jk}$ determines the amplitude of the corresponding pair-creation term between modes $j$ and $k$. Thus, the pairing matrix $\mathrm Z$ provides a compact representation of the photon-pair structure and uniquely specifies the pure Gaussian state.

\begin{figure}[t]
\centering\includegraphics[width=\columnwidth]{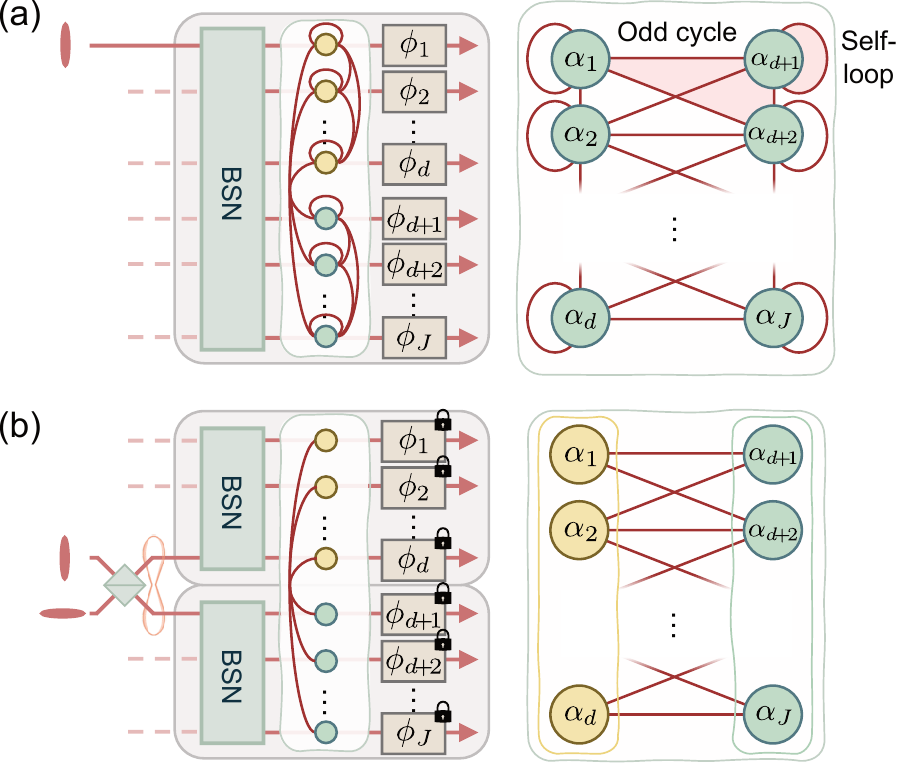}
   \caption{
Example graph structures and optical implementations for a system with $J$ sensing nodes.
(a) An SMSV is distributed over the sensing nodes through a beam-splitter network (BSN). The corresponding graph contains self-loops and odd cycles, which violate the privacy condition.
(b) A TMSV is distributed over the two sensing-node sectors through separate BSNs. The resulting graph contains no self-loops and connects only the two opposite sectors, forming the bipartite structure required by the privacy condition.
}
\label{fig2}
\end{figure}

Then, a local phase encoding $\hat U_{\bm{\phi}}$ acts on the mode creation operator as $\hat U_{\bm{\phi}}\hat a_j^\dagger \hat U_{\bm{\phi}}^\dagger=e^{-i\phi_j}\hat a_j^\dagger.$ Accordingly, defining $\Phi(\boldsymbol \phi):=\operatorname{diag}(e^{-i\phi_1},\ldots,e^{-i\phi_J}),$ the pure Gaussian state in Eq.~\eqref{eq:stateafterUK} after phase encoding can be written as $|\psi(\boldsymbol \phi)\rangle=\mathcal{N}_Z \exp\!\left(\frac{1}{2} \hat{\boldsymbol{a}}^{\dagger\mathsf{T}} \Phi(\boldsymbol \phi) \mathrm Z \Phi(\boldsymbol \phi) \hat{\boldsymbol{a}}^\dagger\right)|0\rangle.$ We define the phase dependent pairing matrix as $\mathrm Z(\bm \phi):= \Phi(\boldsymbol \phi) \mathrm Z \Phi(\boldsymbol \phi),$ which fully describes the phase encoded pure Gaussian state. Therefore, the privacy condition in \eqref{eq:priv} can then be expressed directly in terms of $\mathrm Z(\bm \phi)$ (see SM Sec.~B for further details)
\begin{equation}
    \sum_{j=1}^{J} \alpha_j \frac{\partial \mathrm Z (\boldsymbol{\phi})}{\partial \phi_j} = 0.
    \label{privZ}
\end{equation}
Equation \eqref{privZ} is further simplified as $\mathrm A_{\bm\alpha} \mathrm Z + \mathrm Z \mathrm A_{\bm\alpha}=0,$ where $\mathrm A_{\bm\alpha}:=\operatorname{diag}(\alpha_1,\ldots,\alpha_J).$ Taking the $(j,k)$ matrix element, the pairing matrix satisfies the privacy condition
\begin{equation}
\begin{cases}
(\alpha_j+\alpha_k) Z_{jk}=0, \quad j\ne k\\
Z_{jj}=0,
\end{cases}
\label{eqz}
\end{equation}
for all $j,k$.
Thus, an off-diagonal element $Z_{jk}$ can be nonzero only when $\alpha_j+\alpha_k=0$, while $Z_{jj}=0$ excludes self-loops. In the corresponding graph representation, this means that edges can connect only nodes associated with opposite signs of $\alpha_j$, in other words, odd cycles and self-loops are not allowed. This simple selection rule naturally motivates a graph-theoretic representation of the pairing matrix $\mathrm{Z}$.

Next, we characterize the general structure of Gaussian states that satisfy the privacy condition. For a pairing matrix $\mathrm Z$ satisfying Eq.~\eqref{eqz}, there exists a permutation matrix $\mathrm P$ that groups the modes according to the magnitude of $\alpha_\ell$. Modes associated with the same magnitude $\alpha_\ell$ form the $\ell$-th block, which is further divided into the $+\alpha_\ell$ and $-\alpha_\ell$ sectors, as illustrated in Fig.~\ref{fig1}(b). 

After this permutation, the pairing matrix takes the block-diagonal form
\begin{equation}
\mathrm P\mathrm Z\mathrm P^{\mathsf T}
=
\bigoplus_{\ell=1}^{L}\mathrm Z_\ell,
\qquad
\mathrm Z_\ell
:=
\begin{pmatrix}
0 & \mathrm E_\ell\\
\mathrm E_\ell^{\mathsf T} & 0
\end{pmatrix},
\label{eq:block_form}
\end{equation}
where $\mathrm E_\ell$ describes the edges between the $+\alpha_\ell$ and $-\alpha_\ell$ sectors within the $\ell$-th block. 
In particular, each element $(\mathrm E_\ell)_{jk}$ gives the pairing amplitude between the $j$-th mode in the $+\alpha_\ell$ sector and the $k$-th mode in the $-\alpha_\ell$ sector. Thus, $\mathrm E_\ell$ contains all physically allowed pairwise connections within the $\ell$-th block.
Accordingly,
$\mathrm E_\ell\in\mathbb C^{d_\ell^{+}\times d_\ell^{-}}$, where $d_\ell^{+}$ and $d_\ell^{-}$ denote the numbers of modes in the $+\alpha_\ell$ and $-\alpha_\ell$ sectors, respectively. The total number of modes in the $\ell$-th block is therefore $d_\ell=d_\ell^{+}+d_\ell^{-}$, with $\sum_{\ell=1}^{L}d_\ell=J$. The off-diagonal form of $\mathrm Z_\ell$ directly reflects the privacy condition: pairings are allowed only between the $+\alpha_\ell$ and $-\alpha_\ell$ sectors, while pairings within the same sector are forbidden. Moreover, distinct $\ell$-blocks are completely disconnected and can therefore be analyzed independently. Figure~\ref{fig1}(b) provides a graphical interpretation of this structure. Modes with the same magnitude $\alpha_\ell$ form an independent $\ell$-block, and the enlarged $\ell=2$ example explicitly illustrates the allowed connectivity between its two opposite sectors.

We next identify the elementary Gaussian states associated with each privacy-preserving block $\mathrm Z_\ell$. To this end, we perform the singular-value decomposition~\cite{golub65}
\begin{equation}
\mathrm E_\ell
=
\mathrm W_\ell
\Lambda_\ell
\mathrm V_\ell^\dagger,
\end{equation}
where $\mathrm W_\ell$ and $\mathrm V_\ell$ are unitary matrices. We denote the nonzero singular values of $\mathrm E_\ell$ by $\lambda_\mu^{(\ell)}$, with $\mu=1,\ldots,q_\ell$ and $q_\ell=\operatorname{rank}(\mathrm E_\ell)$. Substituting this decomposition into $\mathrm{Z}_l$ gives
\begin{equation}
\begin{pmatrix}
\mathrm W_\ell^\dagger & 0\\
0 & \mathrm V_\ell^{\mathsf T}
\end{pmatrix}\mathrm
Z_{\ell}
\begin{pmatrix}
\mathrm W_\ell^{*} & 0\\
0 & \mathrm V_\ell
\end{pmatrix}
=
\begin{pmatrix}
0 & \Lambda_\ell\\
\Lambda_\ell^{\mathsf T} & 0
\end{pmatrix}
.
\end{equation}
The outer unitary matrices correspond to passive interferometers acting independently within the two $\alpha$-sectors. Therefore, the central matrix determines the elementary Gaussian states associated with the block. After a further permutation of the modes, it can be written as  
\begin{equation}
\begin{aligned}
\mathrm P'
\begin{pmatrix}
0 & \Lambda_\ell \\
\Lambda_\ell^{\mathsf T} & 0
\end{pmatrix}
\mathrm P'^{\mathsf T}
&=
\bigoplus_{\mu=1}^{q_{\ell}}
\begin{pmatrix}
0 & \lambda^{(\ell)}_{\mu} \\
\lambda^{(\ell)}_{\mu} & 0
\end{pmatrix}
\oplus \bold 0_{d_\ell-2q_\ell}.
\end{aligned}
\end{equation}
Thus, the central matrix can be decomposed into independent $2\times2$ blocks. Writing $\lambda_\mu^{(\ell)}=\tanh r_\mu^{(\ell)}$, each such block takes the form
\begin{equation}
\mathrm Z_{\mathrm{TMSV}}(r_\mu^{(\ell)})
=
\begin{pmatrix}
0 & \tanh r_\mu^{(\ell)}\\
\tanh r_\mu^{(\ell)} & 0
\end{pmatrix}
=\begin{pmatrix}
0 & \lambda_\mu^{(\ell)}\\
\lambda_\mu^{(\ell)} & 0
\end{pmatrix}.
\label{ZTMSV}
\end{equation}
Substituting this pairing matrix into Eq.~\eqref{eq:stateafterUK} gives
\begin{align}
|\psi(r_\mu^{(\ell)})\rangle
&=
\frac{1}{\cosh r_\mu^{(\ell)}}
\sum_{n=0}^{\infty}
(\tanh r_\mu^{(\ell)})^n|n,n\rangle
=
|\mathrm{TMSV}(r_\mu^{(\ell)})\rangle,
\end{align}
showing that each $2\times2$ block corresponds to a two-mode squeezed vacuum (TMSV) state. Therefore, each nonzero singular value $\lambda_\mu^{(\ell)}$ corresponds to an independent TMSV component, while the remaining zero blocks correspond to vacuum modes.

\begin{figure}[t]
\centering\includegraphics[width=\columnwidth]{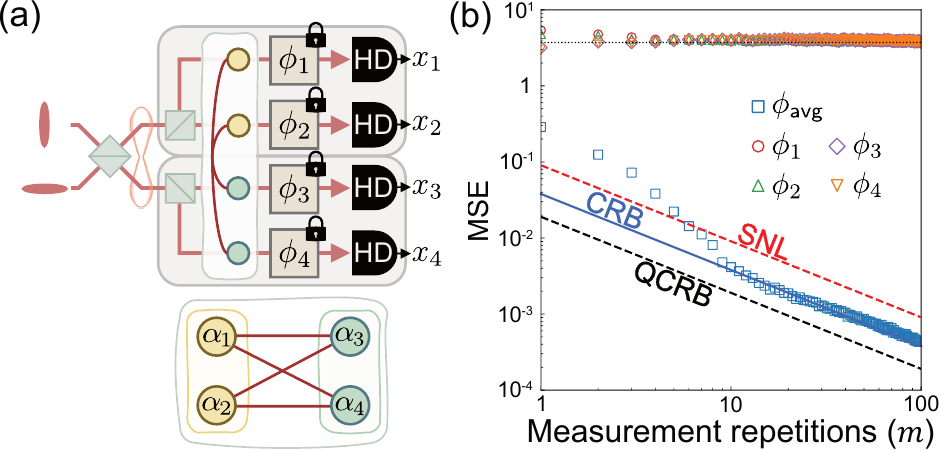}
    \caption{(a) Schematic of the sensitivity-optimal scheme for 4-mode private DQS, where the target weight is $\mathbf{w}=(1,1,1,1)/4$. A TMSV is distributed through BS to local phase shifts, and each node performs local homodyne measurement (HD).
(b) MSEs of maximum-likelihood estimates for both the target and individual local parameters, with $r=1$ and $(\phi_1,\phi_2,\phi_3,\phi_4)=(-2,-1,1,2)$, evaluated over 3000 independent estimations. The target MSE approaches to CRB (blue solid line) with the number of measurement repetitions, whereas the local MSEs remain near the uniform-random-guessing benchmark (dotted line). The shot-noise limit (SNL, red dashed line) is obtained using a coherent state probe with the same mean photon number.
}
\label{fig3}
    \end{figure}

We thus obtain the decomposition
\begin{equation}
|\psi_{\rm priv}\rangle
=
\hat U_{\mathrm K}
\left[
\bigotimes_{\ell,\mu}
|\mathrm{TMSV}(r^{(\ell)}_{\mu})\rangle
\otimes
|\bold 0\rangle
\right]
,
\label{purepriv}
\end{equation}
where $\hat U_{\mathrm K}$ is a passive interferometer that acts independently within each $\alpha$-sector and does not mix modes belonging to opposite sectors. This decomposition shows that every pure Gaussian state satisfying the perfect local privacy condition can be constructed from TMSV states and vacuum modes, followed by passive linear-optical transformations. As illustrated in Fig.~\ref{fig2} (b), the corresponding graph contains only the connectivity structures allowed by the privacy condition. Furthermore, in SM Sec.~H, we extend this characterization to general Gaussian states and completely classify all private states, including mixed states.
    
Figure~\ref{fig2} illustrates this structural condition using two representative Gaussian network configurations. In Fig.~\ref{fig2}(a), a single-mode squeezed vacuum (SMSV) is distributed over the sensing nodes through a beam-splitter network (BSN), which is known to provide the optimal sensitivity among Gaussian probe states for distributed phase sensing \cite{oh20, oh22, matsu19,gessn20,gatto19}. However, the resulting graph generally contains self-loops and odd cycles, which are incompatible with the privacy condition. In contrast, Fig.~\ref{fig2}(b) shows a TMSV distributed through two independent BSNs acting on the two mode sectors. The resulting graph contains no self-loops and connects only opposite sectors, giving the bipartite structure required for perfect local privacy. This example provides a direct physical interpretation of the general decomposition in Eq.~\eqref{purepriv}. See details in SM Sec.~C.

We next optimize the sensitivity of Gaussian probes satisfying perfect local privacy for estimating the weighted global phase 
$\phi=\sum_{j=1}^{J}w_j\phi_j$, with positive weights under a fixed total signal photon number
$N_s=\langle\psi_{\rm priv}|\hat N_s|\psi_{\rm priv}\rangle$, where
$\hat N_s=\sum_{j=1}^{J}\hat n_j$ and $\hat n_j=\hat a_j^\dagger\hat a_j$. For a pure probe, the weak quantum Cram\'er--Rao bound (QCRB) gives \cite{gessn18, paris09,yoo26}
\begin{equation}
    \Delta^2\phi
    \geq
    \bm w^{\mathsf T}H^+\bm w
    \geq
    \frac{w_{\rm tot}^2}
    {4\,\mathrm{Var}(\hat N_s)_{|\psi_{\rm priv}\rangle}},
    \label{QFI_bound}
\end{equation}
where $w_{\rm tot}=\sum_{j=1}^{J}w_j$.
Thus, for a fixed mean photon number $N_s$, minimizing the estimation variance is equivalent to maximizing the photon-number variance of the probe under perfect local privacy. Within the decomposition in Eq.~\eqref{purepriv}, the photon-number variance is maximized by concentrating all squeezing into a single TMSV pair, while $\hat U_{\rm K}$ can be freely chosen within the privacy-preserving block structure. This gives the optimal bound 
\begin{equation}
\Delta^2\phi
\geq
\frac{w_{\rm tot}^2}
{4N_s(N_s+2)}.
\label{QCRB}
\end{equation}
Labeling the single TMSV pair as $(\ell,\mu)=(1,1)$, the corresponding probe can be written as
\begin{equation}
    |\psi_{\rm opt}\rangle
    =
    \hat U_{\rm K}
    \left[
    |\mathrm{TMSV}(r^{(1)}_{1})\rangle
    \otimes
    |\boldsymbol{0}\rangle^{\otimes(J-2)}
    \right].
\end{equation}
A direct evaluation of $\boldsymbol w^{\mathsf T}H^+\boldsymbol w$ for this state shows that the first inequality in Eq.~\eqref{QFI_bound} is also saturated (See SM Sec.~D). Therefore, $|\psi_{\rm opt}\rangle$ attains the bound in Eq.~\eqref{QCRB} and is optimal within the class of Gaussian probes satisfying perfect local privacy. The detailed derivation, including the case of signed weights, is provided in SM Sec.~D.
In particular, for average-phase sensing, $w_j=1/J$, choosing $\hat U_{\rm K}$ to distribute the two TMSV arms uniformly over the $J$ sensing modes gives
$\bar n=\langle\psi_{\rm opt}|\hat n_j|\psi_{\rm opt}\rangle=N_s/J$ for every $j$.
Equation~\eqref{QCRB} then yields
$\Delta^2\phi_{\rm avg}\geq 1/[4J\bar n(J\bar n+2)]$,
demonstrating HS with respect to both the mode number $J$ and the photon number $\bar n$.

Next, we numerically demonstrate the optimal private DQS scheme using a TMSV probe, BSN, and local homodyne measurements. For practical DQS, local measurements at individual sensing nodes are preferable to collective measurements that require recombining the distributed optical modes~\cite{guo20,alushi26}. We therefore adopt local homodyne detection at each node and use maximum-likelihood estimation (MLE) to reconstruct the encoded phases from the measurement outcomes \cite{braunstein92,paris09}.

At each sensing node, as shown in Fig.~\ref{fig3}(a), we measure the quadrature $\hat x_j=(e^{-i\theta_j}\hat a_j
+e^{i\theta_j}\hat a_j^\dagger)/\sqrt{2}$ 
where $\theta_j$ is the local-oscillator phase. Collecting the outcomes from all $J$ nodes gives
$\bm x=(x_1,\ldots,x_J)^{\mathsf T}$.
For the configuration in Fig.~\ref{fig3}(a), the homodyne outcomes are characterized by the covariance matrix
$\mathrm C_{ij}:=\langle\Delta\hat x_i\Delta\hat x_j\rangle$, which takes the form (see SM Sec.~E)
\begin{equation}
\mathrm C_{ij}
=
\begin{cases}
\frac12\delta_{ij}
+
\frac{N_s}{w_{\rm tot}}
\sqrt{w_iw_j}
\cos\!\left(
\varphi_i-\varphi_j
\right),
& \alpha_i=\alpha_j,\\[1ex]
\frac{\sqrt{N_s(N_s+2)}}{w_{\rm tot}}
\sqrt{w_iw_j}
\cos\!\left(
\varphi_i+\varphi_j
\right),
& \alpha_i=-\alpha_j.
\end{cases}
\end{equation}
where $\varphi_j:=\phi_j-\theta_j$ denotes the relative phase between the encoded phase and the local oscillator at node $j$. The two cases correspond to modes in the same and opposite $\alpha$-sectors, respectively. In particular, the local variance
$\mathrm C_{jj}=1/2+(N_sw_j)/w_{\rm tot}$
is independent of $\phi_j$.

From the covariance matrix $\mathrm C$, the CFIM is given by~\cite{oh20}
\begin{equation}
\mathbf F_{ij}=
\frac12
\operatorname{Tr}
\left[
\mathrm C^{-1}
\frac{\partial \mathrm C}{\partial\phi_i}
\mathrm C^{-1}
\frac{\partial \mathrm C}{\partial\phi_j}
\right].
\end{equation}
We choose the local-oscillator phases $\boldsymbol{\theta}$ to minimize the Cram\'er--Rao bound (CRB) $ \Delta^2\phi \ge\mathbf w^{\mathsf T} \mathbf F^+ \mathbf w$.
The homodyne outcomes then follow the zero-mean multivariate Gaussian distribution
$p(\boldsymbol{x}|\boldsymbol{\phi})=\mathcal N
\left(
\boldsymbol{x};
\boldsymbol{0},
\mathrm C_{\boldsymbol{\theta}}(\boldsymbol{\phi})
\right).$
For $m$ measurement repetitions, we generate
$\bm x_m=\{\bm x^{(1)},\ldots,\bm x^{(m)}\}$,
where each outcome
$\bm x^{(s)}=(x_1^{(s)},\ldots,x_J^{(s)})^{\mathsf T}$
is randomly sampled from
$p(\boldsymbol{x}|\boldsymbol{\phi}_{\rm true})$.
In the simulation, we set
$\boldsymbol{\phi}_{\rm true}=(-2,-1,1,2)^{\mathsf T}$.
The phases are estimated by MLE,
$\boldsymbol{\phi}_{\rm est}(\bm x_m)
=
\arg\max_{\boldsymbol{\phi}}
\prod_{s=1}^{m}
p(\bm x^{(s)}|\boldsymbol{\phi})$ \cite{pezze25},
and the corresponding estimate of the authorized global phase is obtained as
$\phi_{\rm est}=\mathbf w^{\mathsf T}
\boldsymbol{\phi}_{\rm est}$.

As shown in Fig.~\ref{fig3}(b), for $J=4$ and
$\mathbf w=(1,1,1,1)^\mathsf T/4$, the MSE of the global-phase estimate decreases with the number of measurement repetitions and approaches the CRB. For each value of $m$, the MSE is evaluated over 3000 independent estimation trials. Using the same homodyne data, we also estimate the individual local phases by MLE. In contrast to the global phase, the corresponding local MSEs remain nearly unchanged with increasing $m$, reflecting the perfect local privacy of the probe. For comparison, the SNL is obtained using the coherent state probe
$|\sqrt{N_s/4}\rangle^{\otimes4}$
with the same total mean photon number $N_s$, giving
$\Delta^2\phi_{\rm SNL}=1/(4mN_s)$~\cite{oh20}. The optimal private Gaussian probe surpasses this SNL while keeping the individual local phases inaccessible.

These results demonstrate that local homodyne measurements together with MLE can simultaneously achieve quantum-enhanced sensitivity and perfect local privacy. We finally note that the QCRB in Eq.~\eqref{QCRB} can, in principle, be saturated by a collective CV-Bell measurement~\cite{de26}. Such a measurement requires recombining the distributed optical modes and is therefore less suitable for DQS. In contrast, the local homodyne scheme avoids mode recombination while retaining HS up to a constant factor (see SM Sec.~E).

We further analyze the robustness of the local privacy condition under phase-independent quantum channels. In realistic optical networks, quantum noise channels, particularly optical loss, may degrade the sensing performance \cite{demkowicz12,escher11,nehra24}. Consider a quantum channel $\mathcal E$ that is independent of the encoded phases and acts on the encoded probe state $\hat\rho_{\boldsymbol\phi}$, such that $\hat\sigma_{\boldsymbol\phi}
:=\mathcal E(\hat\rho_{\boldsymbol\phi})$.
Defining the directional derivative along the privacy direction $\boldsymbol\alpha$ as  $\partial_{\boldsymbol\alpha}:=\sum^J_j \alpha_j\frac{\partial}{\partial\phi_j}$, the privacy condition in Eq.~\eqref{eq:priv} transforms as
\begin{equation}
\begin{aligned}
\partial_{\boldsymbol\alpha}
\hat\sigma_{\boldsymbol\phi}
&=
\partial_{\boldsymbol\alpha}
\mathcal E(\hat\rho_{\boldsymbol\phi})
\\
&=
\mathcal E
\left(
\partial_{\boldsymbol\alpha}
\hat\rho_{\boldsymbol\phi}
\right)
=
0.
\end{aligned}
\label{eq:privacy_channel}
\end{equation}
The second equality follows from the linearity and trace-norm continuity of quantum channels \cite{holevo08,shirokov08,jokinen24}, while the final equality follows from the original privacy condition
$\partial_{\boldsymbol\alpha}\hat\rho_{\boldsymbol\phi}=0$.

Thus, a phase-independent quantum channel cannot restore information along a parameter direction that is already hidden in the input state. Consequently, the perfect local privacy condition is preserved under arbitrary phase-independent quantum channels, including optical loss, phase-insensitive gain, thermal noise, and fixed mode-mixing processes. Such channels may reduce the information available about the authorized collective parameter and therefore degrade the sensitivity. Nevertheless, whenever the authorized collective parameter remains estimable after the channel, perfect local privacy is preserved.

In summary, we have established a complete characterization of Gaussian states satisfying perfect local privacy in DQS. We identified their general graph and state structures, derived the optimal sensitivity under the privacy constraint, and found the corresponding optimal Gaussian probe. We further showed that Heisenberg scaling can be retained with local homodyne measurements and that the local privacy condition is preserved under arbitrary parameter-independent quantum channels, including optical loss. Our results provide a general framework for designing Gaussian quantum sensing networks that simultaneously achieve quantum-enhanced sensitivity and perfect local privacy.

\begin{acknowledgments}
This work was supported by the National Research Foundation of Korea (NRF) grants funded by the Korean government (MSIT) (RS-2024-00352325, RS-2023-NR068116), the Global Learning \& Academic Research Institution for Master’s·PhD Students and Postdocs (LAMP) Program of the NRF funded by the Ministry of Education (RS-2024-00442483), and the Institute for Information \& Communications Technology Planning \& Evaluation (IITP) grants funded by the Korean government (MSIT) (RS-2022-II221029, RS-2025-25464481).
\end{acknowledgments}

\nocite{*}
\bibliography{Reference}

\end{document}